\documentclass[prd,twocolumn,floatfix,preprintnumbers,letterpaper]{revtex4}
\usepackage{graphicx}
\usepackage{amsmath,amssymb}
\input{epsf}
\usepackage{epsf}
\usepackage{xcolor}
\newcommand {\ga} {\ {\raise-.5ex\hbox{$\buildrel>\over\sim$}}\ }
\newcommand {\la} {\ {\raise-.5ex\hbox{$\buildrel<\over\sim$}}\ }

\begin{document}

\def\be{\begin{equation}}
\def\ee{\end{equation}}

\title{Evolution of decaying particles and decay products in various scenarios for the future expansion of the universe}
\author{Cameron E. Norton$^{1,2}$ and Robert J. Scherrer$^{1}$}
\affiliation{$^1$Department of Physics \& Astronomy, Vanderbilt University,
Nashville, TN~~37235}
\affiliation{$^2$Department of Physics, New York University, New York, NY~~
10003}

\begin{abstract}
We examine nonrelativistic particles that decay into relativistic products 
in big rip, little rip, and pseudo-rip models for the future evolution of the universe.
In contrast to decays
that occur in standard $\Lambda$CDM, the evolution of the ratio $r$
of the energy density of the relativistic decay products to the energy density of the initially decaying particles
can decrease with time in all of these models.  In big rip and little rip models, $r$ always
goes to zero asymptotically, while this ratio evolves to infinity or a constant in pseudo-rip models.
\end{abstract}

\maketitle

The evolution of decaying particles in the context of the expanding universe has long been a topic of interest
\cite{DKT1,DKT2,Lindley,Wein,TSK,DK,ENS,ST1,ST2,ST3,LMK1,LMK3}. 
In the early universe, when the expansion is dominated by either radiation or nonrelativistic matter,
exponential decay always leads to the same general result:  the disappearance of the
initial decaying particles and the production of the corresponding decay products, with the density
of the latter always eventually dominating the former.  However, Ref. \cite{KS}
provided an interesting caveat to this result.  If the decaying particles are nonrelativistic,
with density $\rho_M$,
and the decay products are relativistic, with density $\rho_R$, then in
a $\Lambda$CDM universe, the ratio $r$ of the density of the decay products to the density of
the decaying particles, 
\begin{equation}
r \equiv \rho_R/\rho_M,
\end{equation}
need not asymptotically approach infinity, as it always does in a radiation or matter dominated expansion.
Instead, for a sufficiently long decay lifetime, this ratio approaches a constant value.  When this
constant is less than 1, the energy density of the decay products never dominates the energy density of
the decaying particles.
Here we extend this work to expansion laws corresponding to big rip, little rip, and pseudo-rip models
and uncover similarly unusual behavior:  in all three models, the
value of $r$ can decrease with time, and in big rip and little rip models, $r$
always asymptotically approaches zero.

For a nonrelativistic component with density $\rho_M$ decaying with lifetime $\tau$
into a relativistic component with density $\rho_R$,
the equations governing the evolution of the decaying particle and its decay products are \cite{ST1}
\begin{eqnarray}
\label{decay1}
\frac{d\rho_M}{dt} &=& -3H \rho_M - \rho_M/\tau,\\
\label{decay2}
\frac{d\rho_R}{dt} &=& -4H \rho_R + \rho_M/\tau,
\end{eqnarray}
where $H$ is the time-dependent Hubble parameter:
\begin{equation}
H \equiv \frac{\dot a}{a} = \left(\frac{8 \pi G \rho}{3}\right)^{1/2},
\end{equation}
with $\rho$ being the total energy density and $a$ the scale factor,
and we assume a flat universe throughout.  (For other types of energy exchange, see Ref. \cite{barrow}).

Eqs. (\ref{decay1})$-$(\ref{decay2}) can be combined
to yield an equation for $r$:
\begin{equation}
\label{r}
\frac{dr}{dt} = \frac{1}{\tau} + \left(\frac{1}{\tau} - H\right) r.
\end{equation}
When the universe is dominated by matter or radiation, $H$ decreases with time,
so we necessarily have 
$r \rightarrow \infty$ as $t \rightarrow \infty$, as expected.

However,
when the universe enters a vacuum-energy dominated state,
$H$ approaches a constant
value, $H_{\Lambda}$, given by
\begin{equation}
H_{\Lambda} = \left(\frac{8 \pi G \rho_{\Lambda}}{3}\right)^{1/2},
\end{equation}
where $\rho_\Lambda$ is the (constant) vacuum energy density.
Defining the time $t_{\Lambda}  \equiv 1/H_\Lambda$ and substituting  
$H =  1/t_\Lambda$ into equation (\ref{r}),
this equation can be solved analytically to yield \cite{KS}
\begin{equation}
\label{solution}
  r = \left(\frac{t_\Lambda}{t_\Lambda - \tau}\right)\left[\exp\left(\left[\frac{1}{\tau}
- \frac{1}{t_\Lambda}\right]t\right) - 1\right],
\end{equation}
where we have taken $r = 0$ at $t=0$.

As noted in Ref. \cite{KS}, Eq. (\ref{solution}) corresponds to two very different
types of evolution, depending on the ratio of $\tau$ to $t_\Lambda$.
For lifetimes short compared to $t_\Lambda$, i.e., when $\tau < t_\Lambda$,
Eq. (\ref{solution}) gives $r \rightarrow \infty$, just
as in the case for decays during a radiation or matter dominated expansion phase.
In this case, the energy density of the decaying particles becomes infinitesimally small
compared to the energy density of the decay products.  However, when $\tau > t_\Lambda$,
the value of $r$ in equation (\ref{solution}) asymptotically approaches a constant, so that
the density of the decaying nonrelativistic particles never disappears relative to the decay-produced
radiation (although both go to zero in the limit of large $t$).  Furthermore, when
$\tau > 2 t_\Lambda$, the asymptotic ratio of the density of the decay-produced radiation to the 
density of the decaying particles never exceeds 1, so that the decay products never even dominate
the decaying particles.

This result may seem counterintuitive, but Ref. \cite{KS} provides
a simple explanation.  The radiation redshifts relative to matter as one extra power of the scale
factor, which corresponds, in a vacuum-dominated universe,
to an exponential function of time.  Then in calculating the ratio of decay-produced radiation to decaying matter,
this exponential factor cancels the exponential decay of the matter, resulting in a constant final
ratio of matter to radiation.

Given this unusual behavior, we are motivated to consider more extreme expansion laws that have been
proposed for the future evolution
of the universe, all of which involve values for $H$ that increase with time.  While the current expansion
is dominated by a combination of dark matter and dark energy, we will be interested in a future epoch in which
the dark energy is completely dominant and the dark matter can be neglected.  Our results will then be equally valid
regardless of whether it is the dark matter itself that is decaying (as was assumed in Ref. \cite{KS}), or some
other nonrelativistic component.

The three models we consider are the big rip model \cite{bigrip1,bigrip2},
the little rip model \cite{littlerip1,littlerip2}, and the pseudo-rip
model \cite{pseudorip}. These models are all characterized by a value of $H$
that increases with time, but with different final outcomes.  In the big rip,
the value of $H$ goes to infinity at a finite time, resulting in a future
singularity.  In the little rip, $H$ increases monotonically and becomes
arbitrarily large, but it never becomes infinite at a finite time.  Finally,
in the pseudo-rip, $H$ increases monotically but asymptotically approaches
a constant value.

Since the time derivative of $H$ is given by
\begin{equation}
\dot H = - \frac{1}{2} \rho (1+w),
\end{equation}
all three of these types of behavior require a dark energy component
with $w < -1$, which violates the weak energy condition (see Ref. \cite{Carroll}
for a detailed discussion).  Hence, from that standpoint all of these models
are {\it a priori} less plausible than either $\Lambda$CDM or dark energy
models with $w > -1$.  However, observations by themselves do not rule
out $w < -1$.  In a flat universe, and using {\it Planck} 2018 data,
weak lensing, baryon acoustic oscillation, and supernova data,
Ref. \cite{Planck} finds
\begin{equation}
w = -1.028 \pm 0.031,
\end{equation}
at the 68\% confidence level.  While the central value of $w < -1$ should not
be taken too seriously, this result illustrates that observations are far
from ruling out these more exotic models.
Furthermore, it is possible for models to approximate $\Lambda$CDM arbitrarily
closely
at present, but then to evolve into any of the three types of future
evolution considered here; a variety of examples of such models are discussed
in Ref. \cite{Astashenok}.  Thus, observational data can never entirely
rule out future big rip, little rip, or pseudo-rip evolution.

Consider first the big rip model \cite{bigrip1,bigrip2}.  Big rip evolution can arise
if the dark energy has a ``phantom" equation of state parameter, $w$ (the ratio
of the dark energy pressure to density), such that $w < -1$.  Then the phantom dark
energy density {\it increases} with $a$, instead of decreasing as is the case for all fluids with $w > -1$.
The evolution of the scale
factor in a universe containing both matter and phantom dark energy is given by \cite{bigrip1}
\begin{equation}
a = a(t_m)[-w + (1+w)(t/t_m)]^{2/3(1+w)},
\end{equation}
where $t_m$ is the time at which the matter and phantom dark energy densities are equal.
Big rip models are characterized by a future singularity:  as $t$ approaches $[w/(1+w)]t_m$, the scale factor
and phantom energy density both go to infinity in a big rip.  It is more natural to express the scale factor in terms
of the time at which this singularity occurs, namely $t_{rip} = [w/(1+w)]t_m$.  Then the Hubble
parameter in the big rip model is
\begin{equation}
\label{Hbigrip}
H = -\frac{2}{3(1+w)(t_{rip} - t)}.
\end{equation}

Using this form for $H$, we have numerically integrated Eq. (\ref{r}) for $w = -1.05$ and
$\tau/t_{rip} = $ 0.1, 0.3, and 1.0; the results are shown in Fig. 1.  The evolution of $r$ for
the big-rip cosmology is strikingly different from its evolution in $\Lambda$CDM.  As is the case
for $\Lambda$CDM, $r$ initially increases, but instead of increasing to arbitrary large values or
asymptotically approaching a constant value (the two possibilities for $\Lambda$CDM), it reaches
a maximum value and then decreases, approaching zero as $t$ reaches $t_{rip}$.  Thus,
in the big-rip cosmology, the decay-produced radiation density is always asymptotically subdominant compared to the density
of the decaying particles, and the ratio between these two quantities becomes arbitrarily small as $t \rightarrow t_{rip}$.
\begin{figure}[t]
\centerline{\epsfxsize=3.7truein\epsfbox{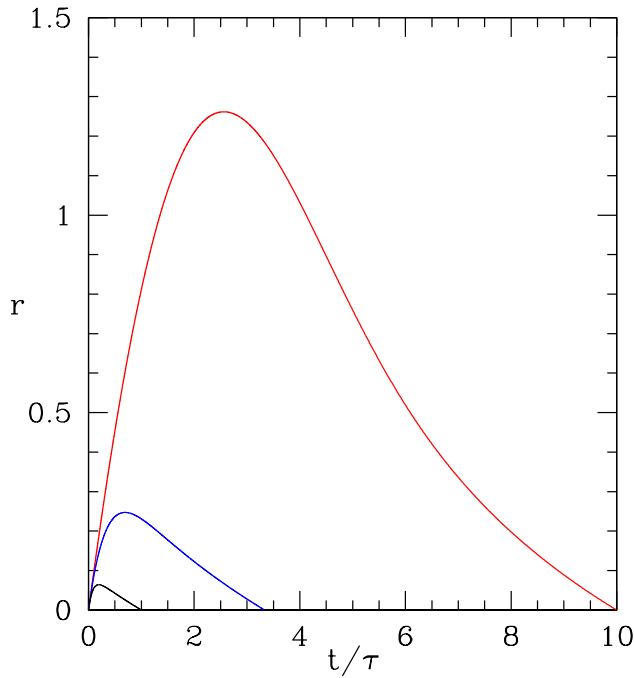}}
\caption{The ratio $r$ of decay-produced relativistic energy density
to the density of decaying nonrelativistic matter
as a function of the time $t$ measured in units of the decaying
particle lifetime $\tau$ for a big-rip cosmology with $w = -1.05$ and a future singularity
at the time $t_{rip}$.  From top to bottom, the curves
correspond to $\tau/t_{rip} = $ 0.1 (red), 0.3 (blue) and 1.0 (black).}
\end{figure}

Big rip evolution represents a rather extreme case, resulting as it does in a future singularity.  A less extreme class
of models, dubbed the ``little rip", occurs when $H \rightarrow \infty$ not at a finite time, but as $t \rightarrow
\infty$ \cite{littlerip1,littlerip2}.  While they do not result in a future singularity, little rip models do lead to the dissolution of all bound structures as
$H$ becomes arbitrarily large. In general, these models correspond to dark energy with a density that
increases with the scale factor, but more slowly than a power law (e.g., logarithmically).
In terms of the time evolution
of the scale factor, any expansion law of the form
\begin{equation}
a = e^{f(t)},
\end{equation}
where $\ddot f > 0$ and $f(t)$ is a nonsingular function of $t$, will correspond to a little rip \cite{littlerip1}.
There is an infinite set of such models, so we will examine one of the simplest, namely
\begin{equation}
a = a_0 e^{(1/2)(t/t_0)^2}.
\end{equation}
Here $a_0$ and $t_0$ are a fiducial scale factor and time, respectively.  Then $H$ is given by
\begin{equation}
\label{littlerip}
H = \frac{t}{t_0^2}.
\end{equation}

Using this value for $H$ in Eq. (\ref{r}), we have generated curves for $r$ as a function
of $t/\tau$, which are shown in Fig. 2 for several values of $\tau/t_0$.
\begin{figure}[t]
\centerline{\epsfxsize=3.7truein\epsfbox{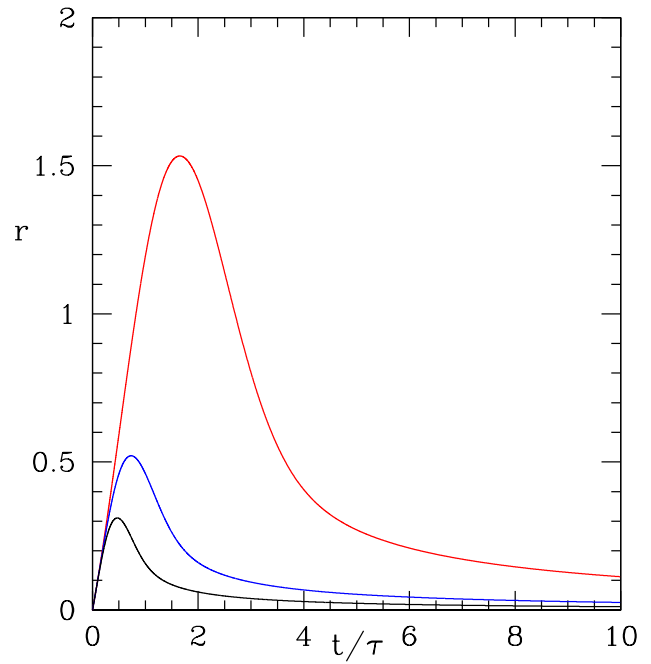}}
\caption{The ratio $r$ of decay-produced relativistic energy density
to the density of decaying nonrelativistic matter
as a function of the time $t$ measured in units of the decaying
particle lifetime $\tau$ for the little rip model
given by Eq. (\ref{littlerip}) with, from top to bottom, $\tau/t_0 = $ 1.0 (red), 2.0 (blue) and 3.0 (black).}
\end{figure}
We see that even for the case of the little rip, the generic late-time evolution is
a ratio of $\rho_R$ to $\rho_M$ that decreases with time.  In this case,
however, the ratio goes to zero as $t \rightarrow \infty$.

Finally, we consider the pseudo-rip \cite{pseudorip}.  In these models, $H$ is always an increasing function of
time, as in the case of the big or little rip models, but it asymptotically approaches a constant, as in
$\Lambda$CDM.  Just as in the case of the little rip, there are an infinite set of such models, so we will
choose a single representative model to examine here, namely
\begin{equation}
\label{pseudorip}
H = H_1 + (H_2 - H_1) \tanh(t/t_0).
\end{equation}
In this model, when $t \ll t_0$, the universe is in a de Sitter phase with constant Hubble parameter, $H=H_1$.
When $t \sim t_0$, $H$ increases
with time,
asymptotically approaching $H_2$, where we take $H_2 > H_1$.
The time $t_0$ simply specifies the characteristic time
at which this
transition takes place.

Taking (somewhat arbitrarily) $t_0 = \tau$, we have integrated Eq. (\ref{r}) for this model;
the corresponding behavior of $r$ as a function of $t$ is shown
in Fig. 3.
\begin{figure}[t]
\centerline{\epsfxsize=3.7truein\epsfbox{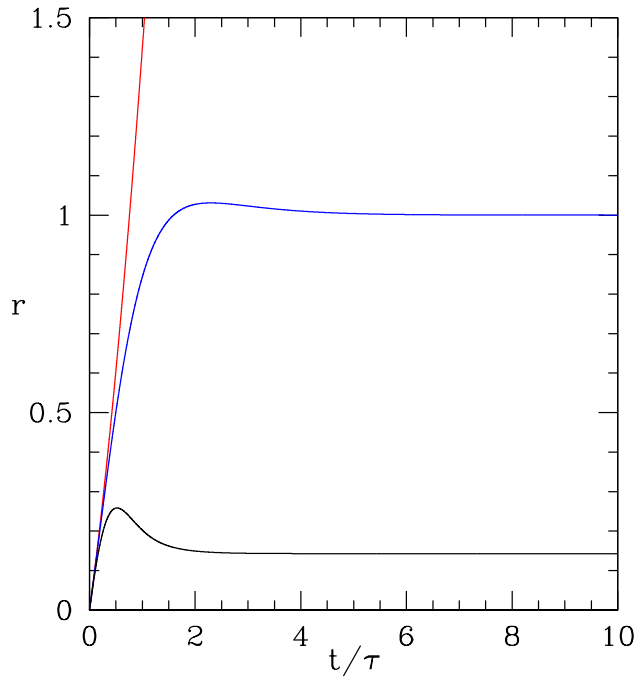}}
\caption{The ratio $r$ of decay-produced relativistic energy density
to the density of decaying nonrelativistic matter
as a function of the time $t$ measured in units of the decaying
particle lifetime $\tau$ for the pseudo-rip model
given by Eq. (\ref{pseudorip}) with $t_0 = \tau$. From top to bottom, the curves
correspond to $H_1 \tau = 0.125$, $H_2 \tau = 0.5$ (red),
$H_1 \tau = 0.5$, $H_2 \tau = 2.0$ (blue)
and $H_1 \tau = 2.0$, $H_2 \tau = 8.0$ (black).}
\end{figure}
The behavior of $r$ in this pseudo-rip model is quite similar to the case of $\Lambda$CDM.
Note that the asymptotic value of $H$ in this model is $H_2$.
When this
asymptotic value satisfies $H_2 \tau < 1$, the value of $r$ evolves to infinity, just as for $\Lambda$CDM.
On the other hand, when the $H_2 \tau > 1$, $r$ evolves to a constant value.  The difference
from $\Lambda$CDM is that $r$ can undergo an earlier phase in which it decreases with time, just as in the case of the big rip
and little rip models; such behavior is impossible for a universe dominated by a cosmological constant or any fluid
with $w > -1$.

As
we have investigated only a single set of example models for each type of future evolution,
it is reasonable to question the extent to which our results are generic.
From Eq. (\ref{r}), we see that at early times, when $(H\tau -1) r < 1$, the value of $r$ necessarily
increases with time, as is evident in all three types of evolution (as well as in standard $\Lambda$CDM).
In the case of the big rip and little rip, $H$ increases at late times to arbitrarily
large values.  When $H$ becomes larger than $1/\tau$, the ratio $r$ begins to decrease.
Finally, when $H \gg 1/\tau$, Eq. (\ref{r}) has an approximate analytic solution, namely
\begin{equation}
\label{solr}
r \approx \frac{1}{H \tau}.
\end{equation}
In big rip models, $H \rightarrow \infty$ at $t_{rip}$, so Eq. (\ref{solr})
tells us that $r \rightarrow 0$ at $t_{rip}$ as well.
Similarly, since $H$ increases to arbitrarily large values in little rip models,
we have $r \rightarrow 0$ in those models.  Eq. (\ref{solr}) has a simple physical
interpretation: if matter is decaying at a rate $1/\tau$, then at a given Hubble time $1/H$, the fraction of the matter that has decayed
into radiation is just $(1/\tau)(1/H)$.

The case of the pseudo-rip more closely resembles $\Lambda$CDM.  Since $H$ asymptotically goes to a constant at late times,
the pseudo-rip evolves asymptotically to $\Lambda$CDM.  Hence, $r$ will be given by Eq. (\ref{solution}) at late times,
with $t_\Lambda$ corresponding to the asymptotic value of $1/H$.  This is apparent in Fig. 3, in that the asymptotic
behavior
of $r$ depends entirely on the value of $H_2 \tau$.  However, the evolution of $H$ at earlier times allows $r$ to evolve
differently from its behavior in $\Lambda$CDM. In particular we can have intervals over which $r$ decreases with
time; such behavior is impossible in $\Lambda$CDM.

In summary, the late-time evolution of the ratio $r$
of the energy density of the relativistic decay products to that of the initially decaying particles depends on the asymptotic
evolution of $H$.  In models for which $H$ is an unbounded increasing function of $t$ (big rip and little rip) the value of $r$ reaches
a maximum and decreases asymptotically to zero, in sharp contrast to the behavior of $r$ in $\Lambda$CDM.  Pseudo-rip models, in
contrast, exhibit a value for $r$ that goes to either infinity or a nonzero constant at late times.

Of course, in all of these models, both the decaying particle energy density and the density of the decay products
rapidly go to zero, and neither has an effect on the overall expansion rate.  What is interesting is the way
that the evolution of these quantities violates our intuition from particle decays in the early universe, when the expansion
is dominated by matter or radiation, and the decaying particle density rapidly becomes subdominant compared to the
density of the decay products.

\acknowledgements 

R.J.S. was supported in part by the Department of Energy (DE-SC0019207).


\begin{thebibliography}{99}

\bibitem{DKT1} D.A. Dicus, E.W. Kolb, and V.L. Teplitz, \prl {\bf 39}, 168 (1977).

\bibitem{DKT2} D.A. Dicus, E.W. Kolb, and V.L. Teplitz, \apj {\bf 221}, 327 (1978).

\bibitem{Lindley} D. Lindley, Mon. Not. R. Astron. Soc. {\bf 188}, P15 (1979).

\bibitem{Wein} S. Weinberg, \prl {\bf 48}, 1303 (1982).

\bibitem{LMK1}  L.M. Krauss, Nucl. Phys. {\bf B227}, 556 (1983).

\bibitem{TSK} M.S. Turner, G. Steigman, and L.M. Krauss,
\prl {\bf 52}, 2090 (1984).

\bibitem{DK} A.G. Doroshkevich and M. Yu. Khlopov, Mon. Not. R. astr. Soc. {\bf 211},
277 (1984).

\bibitem{LMK3}  L.M. Krauss, Gen. Rel. and Grav. {\bf 17}, 89 (1985).

\bibitem{ENS} J. Ellis, D.V. Nanopoulos, and S. Sarkar, Nucl. Phys. B {\bf 259}, 175 (1985).

\bibitem{ST1} R.J. Scherrer and M.S. Turner, \prd {\bf 31}, 681 (1985).

\bibitem{ST2} R.J. Scherrer and M.S. Turner, \apj {\bf 331}, 19 (1988).

\bibitem{ST3} R.J. Scherrer and M.S. Turner, \apj {\bf 331}, 33 (1988).

\bibitem{KS} L.M. Krauss and R.J. Scherrer, \prd {\bf 75}, 083524 (2007).

\bibitem{barrow} J.D. Barrow and T. Clifton, Phys. Rev. D, 73, 103520 (2006).

\bibitem{bigrip1}
R.R. Caldwell,
Phys. Lett. B {\bf 545}, 23 (2002).

\bibitem{bigrip2}
R.R. Caldwell, M. Kamionkowski, and N.N. Weinberg,
\prl {\bf 91}, 071301 (2003).

\bibitem{littlerip1}  P.H. Frampton, K.J. Ludwick, and R.J. Scherrer,
\prd {\bf 84}, 063003 (2011).

\bibitem{littlerip2}  P.H. Frampton, K.J. Ludwick, S. Nojiri, S.D. Odintsov, and R.J. Scherrer,
Phys. Lett. B {\bf 708}, 204 (2012).

\bibitem{pseudorip} P.H. Frampton, K.J. Ludwick, and R.J. Scherrer, \prd {\bf 85},
083001 (2012).

\bibitem{Carroll} S.M. Carroll, M. Hoffman, and M. Trodden, \prd
{\bf 68}, 023509 (2003).

\bibitem{Planck} N. Aghanim, et al., Astron. Astrophys. {\bf 641}, A6 (2020).

\bibitem{Astashenok} A.V. Astashenok, S. Nojiri, S.D. Odintsov, and
R.J. Scherrer, Phys. Lett. B {\bf 713}, 145 (2012).

\end{thebibliography}
\end{document}